\documentstyle[epsf,twocolumn,prl,aps]{revtex}

\begin{document}
\draft

 \twocolumn[\hsize\textwidth\columnwidth\hsize\csname @twocolumnfalse\endcsname

\title{Density Matrix Approach to Local Hilbert Space Reduction}
\author{Chunli Zhang, Eric Jeckelmann, and Steven R.\ White}
\address{ 
Department of Physics and Astronomy,
University of California,
Irvine, CA 92697
}
\date{\today}
\maketitle
\begin{abstract}
\noindent 

We present a density matrix approach for treating systems with a large or
infinite number of degrees of freedom per site with exact
diagonalization or the density matrix renormalization group.
The method is demonstrated on the 1D Holstein model of electrons
coupled to Einstein phonons. In this system, two or three
optimized phonon modes per site give results as accurate as with 10-100 bare
phonon levels per site. 

\end{abstract}

 ]

During the past decade there have been great strides in
the development of numerical techniques for simulating strongly
correlated systems. A significant limitation to many of these
methods---for example, exact diagonalization using the Lanczos
algorithm, or the density matrix renormalization
group\cite{dmrg} (DMRG)---is that they require a finite basis. 
In an electron-phonon lattice model, for example, the number of
phonons is not conserved and the Hilbert space is infinite,
even for a finite number of sites. Of course, the number of
phonons can be artificially constrained, but for strongly
coupled systems the number of phonons needed for an accurate
treatment may be quite large. This often severely constrains the
size of the systems which may be studied.

Here we present a technique for generating a controlled
truncation of the Hilbert space,  which allows the use of a
very small local basis without significant loss of accuracy.
The local basis which is generated can be used in exact
diagonalization, DMRG, or other approaches to allow treatment
of larger systems. The procedure is closely related to 
DMRG, in that the local basis is generated using a density
matrix, but it is simpler to implement. We illustrate the
method on the Holstein model, a model of noninteracting
electrons on a lattice coupled to phonons, with one Einstein 
oscillator on each site. For this model we show that with 
two or three optimized phonon modes per site, we obtain the 
same accuracy as with dozens of unoptimized phonon levels per site. 
For simplicity,
the Hilbert space reduction technique is used here in conjunction
with exact diagonalization, but coupling it to other approaches,
such as DMRG, would allow treatment of larger systems.

The key idea of this approach is identical to the key idea of DMRG\cite{dmrg}:  
in order to eliminate states from a part of a system without loss
of accuracy, one should transform to the basis of eigenvectors of the 
reduced density matrix, and discard states with low probability.
The key difference is that here the subsystem is a single 
site, or a handful of sites, rather than varying fractions of
the entire system.
To be specific,
consider a many-body system divided into unit cells which we
will call ``sites'', such that the
complete Hilbert space of the system is the outer product of the
states of the sites. The number of states per site may be
finite or infinite.  We will consider here only systems with
translational invariance, so that all sites are equivalent. 
Let $i$ label the states of a particular single site, say site 1.
Let $j$ label the combined states of all of the rest of the
sites. Then a wavefunction of the system can be written as
\begin{equation}
|\psi\rangle = \sum_{ij} \psi_{ij} | i \rangle | j \rangle .
\end{equation}
The density matrix for this site when the system is in the state
$|\psi\rangle $ is
\begin{equation}
\rho_{ii'} = \sum_{j} \psi_{ij}\psi^*_{ij} .
\end{equation}
Let $w_\alpha$ be the eigenvalues of $\rho$, and let $v_\alpha$
be the eigenvectors. The $w_\alpha$ are the probabilities of the
states $v_\alpha$. If $w_\alpha$ is negligible, then the
corresponding eigenvector $v_\alpha$ can be discarded from the
basis for the site, without affecting the state $\psi$. If one
wishes to keep a limited number of states $m$ per site, then the best
states to keep are the eigenstates of $\rho$ with largest
eigenvalues\cite{dmrg}. In the case of the Holstein model, we
will show that all but a handful of these eigenstates have
negligible probability.

Usually the target state $\psi$ which one wants to represent is the
ground state.  If one wants a site basis which represents several states,
one can add each state into the density matrix
\begin{equation}
\rho_{ii'} = \sum_\alpha \sum_{j} a_\alpha \psi^\alpha_{ij}{\psi^\alpha_{ij}}^* ,
\end{equation}
where the $a_\alpha$ are weights assigned to each target state,
representing the importance of that state.
Again, the optimal states to keep are the eigenstates of $\rho$. 

Unfortunately, in order to obtain the optimal states, we need
the target state, which we usually we do not know--usually we want 
the optimal states to help get the target state. This problem
can be circumvented in several ways. We illustrate these
approaches in the Holstein model, which has as Hamiltonian
\begin{eqnarray}
H =  \omega \, \sum_\ell b^\dag_\ell b_\ell 
- \gamma \, \sum_\ell \left (b^\dag_\ell + b_\ell \right ) 
n_\ell
\nonumber \\ - t \sum_{\ell\sigma} \left (c^\dag_{\ell+1\sigma}
c_{\ell\sigma} + c^\dag_{\ell\sigma} c_{\ell+1\sigma} \right )  \, ,
\label{eq:ham}
\end{eqnarray}
where $c^\dag_{\ell\sigma}$ and $c_{\ell\sigma}$ are electron creation and
annihilation 
operators, $b^\dag_\ell$ and $b_\ell$ are
phonon creation and annihilation operators,
$n_\ell = c^\dag_{\ell\uparrow} c_{\ell\uparrow}+c^\dag_{\ell\downarrow}
c_{\ell\downarrow}$ and $t$ is the hopping
integral, $\gamma$ is the electron-phonon coupling constant, and
each oscillator has frequency $\omega$.

The first approach is
illustrated in Fig. 1(a). Here one site of the
system (the ``big site'') is allowed to 
have a large number of phonon states $M$,
with $M \sim 10 - 100$. The rest of the sites have a much smaller
number of phonon levels, $m \sim 2 - 3$. A set of Davidson or
Lanczos exact diagonalizations are performed. In the first
diagonalization, all of the phonon states are ``bare'': they are
eigenstates of the single site phonon Hamiltonian, characterized
by the frequency $\omega$. After each diagonalization, the
density matrix for the phonon modes of the
big site is
diagonalized. The most probable $m$ eigenstates
are the new optimal phonon modes. These optimal phonon modes 
are used on all the other sites for the next 
diagonalization. In each diagonalization, the big site
always has a large number of phonon modes, so that it can
generate improved optimal modes for the next iteration. 
The diagonalizations are repeated until the optimal modes have
converged. In the first diagonalization, the optimal modes are
not very accurate because the bare phonon levels used on the
other sites severely truncate the Hilbert space. Convergence
takes only a few diagonalizations, however.

\vbox{
\begin{figure}[!b]
\epsfxsize=2.575 in\centerline{\epsffile{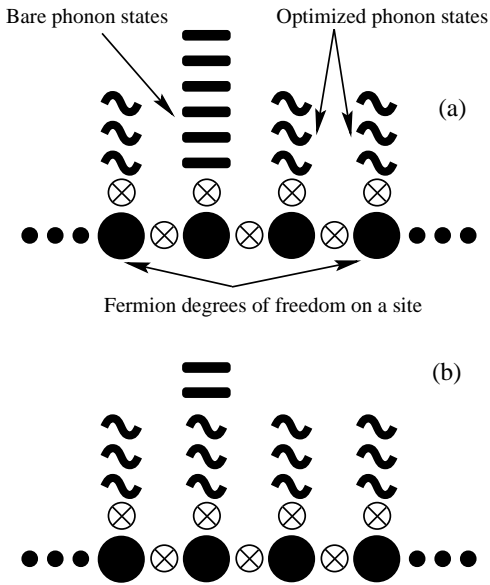}}
\caption{
Algorithms for constructing optimal bases. (a) The
big  site has the complete set of bare phonon levels. (b) The
big site has the optimal levels plus a few bare levels.
}
\end{figure}
}

While this approach is very simple, the large number of states
on the big site limits the size of the system which can be
diagonalized. A more sophisticated approach is illustrated in
Fig. 1(b). Here, the big site has both the optimal
modes and two or three extra levels, rather than  $M$
bare levels. These extra levels allow improvement of the optimal
basis. They are taken from the set of $M$ bare levels but are
explicitly orthogonalized to the current optimal modes. After 
a diagonalization including these levels, a new density matrix
is formed and optimal modes found. These optimal modes can mix in a
little of the extra levels to improve the basis. The next
diagonalization uses different extra levels. One sweep consists
of enough diagonalizations to include all $M$ bare levels as
extra levels. A couple of sweeps are needed to reach full
convergence of the optimal levels. Each diagonalization uses as
the starting wavefunction the converged wavefunction from the last
step. Therefore only two or three Davidson steps are needed for
convergence, rather than dozens.

A further improvement comes from forming the density matrix for
an entire site, including electron degrees of freedom, 
rather than just the phonon levels. This forms
different optimal phonon modes for each of the four electron states of the
site, which reduces the number of states needed for a given
accuracy.

\begin{figure}[ht]
\epsfxsize=3.375 in\centerline{\epsffile{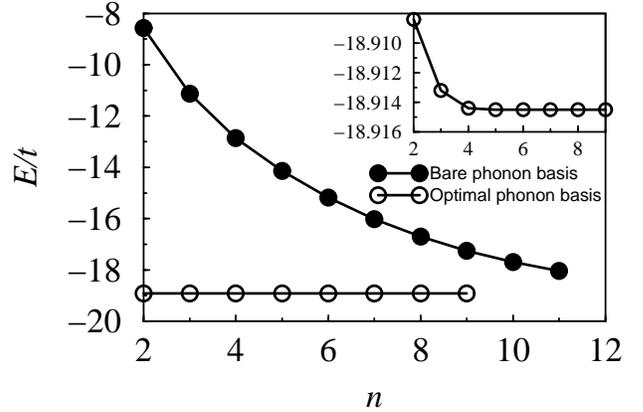}}
\caption{Ground state energy of a 4-site half-filled Holstein system
with $\omega=t$ and $\gamma=1.5t$ as a function of the number of
phonon states kept on each site of the lattice.  
}
\end{figure}

In Fig. 2 we show the ground state energy of a four site system
as a function of the number of phonon modes, both for bare
phonon levels and for optimal modes, where the $m$ optimal modes are
allowed to be different for each electron state of a site. The
energy in these restricted bases form upper bounds to the true
ground state energy of the four site system. The
improvement coming from using optimal modes is remarkable. 
Using only two optimal modes, the energy is accurate to less than 0.1\%,
whereas keeping eleven bare modes (the largest number we could treat), 
the error is greater than 5\%.

The very rapid convergence of the energy as a function of the number of 
optimal phonon modes is due to the very tight distribution of their
probabilities $w_\alpha$. In two special cases ($\gamma = 0$ or $t=0$)
it is possible to show that at most one optimal phonon level 
for each electron state of a site has a non-zero probability.
For arbitrary couplings several optimal phonon levels can have a 
finite probability but only a few of them are significant.
In the ground state of a 6-site Holstein model with $\omega=t$,
we have found that the third and fourth highest eigenvalues $w_\alpha$
(for a given electron state of a site) are smaller than $10^{-3}$
and $10^{-5}$, respectively, for any electron-phonon coupling~$\gamma$.
Therefore, we have never needed more than $m = 2-3$ optimal phonon levels
to get an accurate ground state.

In Fig.~3(a) we show the four most probable phonon states 
expressed in the bare phonon basis 
for the ground state of a 6-site Holstein lattice.
The probability $w_\alpha$ and the occupation of the electronic site 
are also given for each state.
It is clear that one needs at least $\sim 20$ bare phonon levels for an
accurate treatment of this system.

An additional improvement is often possible: the optimal levels
may be transferable from a smaller system to a larger system.
We find that in the Holstein model at half filling, 
the levels obtained from applying this procedure to a two site or four site
system work very well for larger systems. Thus for the larger
system, one needs to do only one diagonalization, and each site 
has only $m$ levels. This would be the simplest way to
incorporate this approach into DMRG: use a small system
diagonalization approach to find optimal levels, and then use
DMRG to treat very long chains. Away from half-filling, because
only certain fillings are possible on small systems, it would
probably be necessary to simultaneously target two or more states with
different fillings in order for the optimal basis to be
transferable to a large system.

The form of the optimal phonon levels can tell us important
information about the system. In Fig.~3(b) we show optimal phonon
wavefunctions as a function of the oscillator position 
$q=b^\dag+b$ for different electron-phonon couplings.
Only the most probable optimized phonon 
level is shown for each possible electron state of a site.
If the optimal states are allowed to be different for each
electron state, we find that 
every optimal state is either an eigenstate of an oscillator
with a shifted equilibrium position or a linear combination of two such 
eigenstates, to surprisingly high accuracy, with overlaps
greater than 96\%.
Unfortunately, we can not use this property 
to calculate a priori a basis of optimal states.

The form of the most important phonon states
shown in Fig.~3(b) can be understood qualitatively in the
weak and strong coupling regimes. The behavior at intermediate
couplings interpolates smoothly between strong and 
weak coupling.
In the weak coupling regime ($\gamma = 0.5t$) 
optimal states are simply
eigenstates of an oscillator with an equilibrium position 
$q\approx 2\gamma/\omega$ as predicted by a mean-

\begin{figure}[ht]
\epsfxsize=3.375 in\centerline{\epsffile{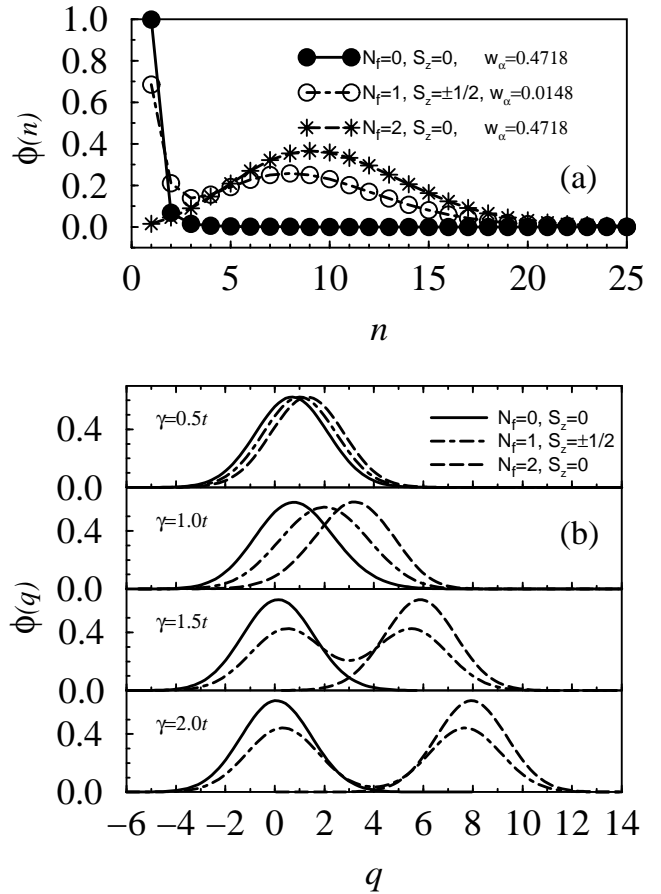}}
\caption{(a)  The first four optimal phonon states expressed
in the bare phonon basis for a 6-site half-filled Holstein system
with $\omega=t$ and $\gamma=1.5t$. 
(b)  Optimal phonon wavefunctions as a function of the oscillator position
$q$ for different electron-phonon couplings $\gamma$ for the same lattice.
}
\end{figure}

\noindent 
field approximation.
Differences between the optimal phonon states for the different 
electronic states of a site are
small compared to their widths, which are determined by phonon fluctuations.
Therefore, electron and phonons are almost independent 
and the lattice relaxation generated by the presence of electrons
is much smaller than quantum lattice fluctuations.

In the strong coupling regime ($\gamma = 2t$) we have found
that the optimal states
for $N_f=0,2$ are very similar to the ground states of oscillators with 
equilibrium positions $q\approx 2N_f \, \gamma/\omega$.
This corresponds to a bipolaronic 
ground state in which electrons are trapped by local lattice distortions
and form pairs localized on a single site, with one pair for
every two sites.
Except for a small shift in the equilibrium positions,
the optimal phonon states for a singly
occupied site are almost the superposition of the optimal phonon states 
for $N_f=0,2$. 
This can be understood as a retardation effect: the
$N_f = 1$ states are intermediate states with low probability;
electrons do not spend enough time in these states for the
phonon states to adapt.
These optimized phonon states for $N_f=1$
are very different from the phonon
states generated by the Lang-Firsov transformation
which shifts each oscillator equilibrium position
by a quantity $2N_f \, \gamma/\omega$.
Therefore, it is not surprising that the standard strong coupling
theory, which is based on the Lang-Firsov transformation,
poorly describes the electronic and dynamical properties
of the Holstein Hamiltonian~\cite{diago}.

Mean-field theory predicts that the ground state
of the half-filled Holstein model is a charge-density-wave
(CDW) state with a dimerized lattice and a gap
at the Fermi surface for any finite electron-phonon coupling.
An interesting question is whether the ground state is modified when quantum
lattice fluctuations are taken into account.
There is strong evidence that the system is dimerized
for arbitrary finite coupling at finite phonon
frequency~\cite{qmc,variational}.
On the other hand, recent results suggest that the gap is destroyed 
by quantum lattice fluctuations in the weak coupling regime~\cite{funcint}.
In the 2-site Holstein model it is known that there is a crossover
from quasi-free electrons 
to a bipolaron at a finite coupling~\cite{diago}.

In Fig. 4 we present several quantities obtained
with our method as a function of the electron-phonon
coupling $\gamma$ for a 6-site system with $\omega=t$.
Our results for the phonon order parameter
$m_p$, defined by
$4 m_p^2 = \langle (q_\ell -q_{\ell+1})^2 \rangle - 2$,
where $q_\ell = b^\dag_\ell + b_\ell$,
are qualitatively similar to the predictions of previous
studies~\cite{qmc,variational}.
Note that with this definition, $m_p=0$ for $\gamma=0$.
We find $m_p \neq 0$ for any finite $\gamma$,
although $m_p$ is smaller than the zero-point lattice fluctuations 
in the weak coupling regime $\gamma \leq 0.8t$.
For $\gamma \geq 1.5t$,
$m_p$ approaches the strong-coupling theory result $m_p=2\gamma/\omega$ 
The electronic static staggered susceptibility, defined as
\begin{equation}
\chi_f =  \sum_\ell (-1)^\ell \langle n_i n_{i+\ell} \rangle  \,,
\end{equation}
indicates the existence of a transition around $\gamma = t$, 
where $\chi_f$ suddenly increases
from the free electron result $\chi_f =1$ and approaches
the  value $\chi_f = 6$ representative of a perfect CDW order in
a 6-site system (Fig. 4).
Finally, in Fig. 4 we also show the next-nearest neighbor pairing 
correlation
$\langle P_\ell P^\dag_{\ell+2} \rangle$, where
 $P_\ell = c_{\ell\uparrow} c_{\ell\downarrow}$. 
This quantity has a peak around the value of $\gamma$
where the dimerization amplitude $m_p$ starts to
dominate quantum lattice fluctuations and
$\chi_f$ goes up rapidly.

In the weak coupling regime we have been able to study larger systems
(with up to 40 sites) with a DMRG method using a different approach
to handle the  phonon
Hilbert space\cite{bosondmrg}.
We have found that despite the peak 
evident in Fig. 4, the
pairing correlations 
$\langle P_\ell P^\dag_{\ell+m} \rangle$
decrease as $1/m^2$, similar to the behavior of free electrons,
even for a coupling as large as $\gamma = 0.8t$.
In the strong coupling regime, 
pairing correlations decay exponentially
because the dimerization opens a gap at the Fermi level~\cite{qmc}. 
Therefore, it is possible that there is a transition
from a metallic to an insulating CDW phase at a finite
electron-

\begin{figure}[ht]
\epsfxsize=3.375 in\centerline{\epsffile{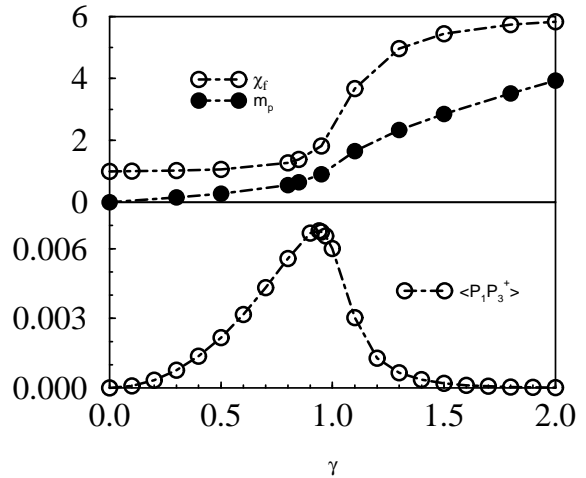}}
\caption{Phonon order parameter $m_p$, staggered static electronic 
susceptibility $\chi_f$ and next-nearest neighbor pairing correlations 
$\langle P_\ell P^\dag_{\ell+2} \rangle$
as a function of the electron-phonon coupling in a 6-site Holstein
lattice with $\omega=t$.}
\end{figure}

\noindent 
phonon coupling~\cite{funcint}.

Obviously, a complete understanding of the Holstein model 
properties requires the study of larger systems.
The use of optimized phonon basis coupled to 
powerful numerical methods, such as DMRG, will enable us to 
perform these calculations.
Similarly, the techniques described in this letter could greatly
improve our capability to perform numerical studies
of other problems which involve an infinite Hilbert space.


We would like to thank J.T. Gammel and Alan Bishop for helpful discussions.  
SRW acknowledges support from the NSF under 
Grant No. DMR-9509945, and from the University of California
through the Campus Laboratory Collaborations Program.
EJ thanks the Swiss National Science Foundation for financial support.
Calculations were performed at the San Diego
Supercomputer Center.

\end{document}